\documentclass[twocolumn,showpacs,preprintnumbers,amsmath,amssymb]{revtex4}

\usepackage{graphicx}
\usepackage{dcolumn}
\usepackage{bm}
\begin{document}
\title{A dynamical equation for the distribution of a scalar advected by turbulence}
\author{Antoine Venaille}
\email{venaille@coriolis-legi.org}
\author{Joel Sommeria}%
\email{sommeria@coriolis-legi.org}
\affiliation{ Coriolis-LEGI 21 rue des martyrs 38000 Grenoble France
}

\date{\today}

\begin{abstract}
A phenomenological model
 for the dissipation of scalar fluctuations due to the straining by the fluid motion is proposed in this letter. An explicit equation is obtained for the time evolution of the probability distribution function of a coarse-grained scalar concentration. The model relies on a self-convolution process. 
We first present this model in the Batchelor regime and then extend empirically our result to the turbulent case. The inclusion of this model in more general transport equations, including spatial gradients, is discussed in relation with 2D turbulence and stratified flows.
\end{abstract}



\maketitle

The turbulent transport of tracers such as temperature or salinity is an important issue for many applications \cite{siggia}.
Available models usually provide a set of closed equations
for the mean quantities and their second moments (variance), and sometimes for the third and the fourth moments \cite{canuto}.
However, it may be useful to model the evolution of the whole
 probability distribution of the scalar. 
This problem has been discussed in the context of reactive flows \cite{pope85}. 
It could be also important to properly model the turbulent mixing of
 water masses in a stably stratified fluid, where sedimentation under gravity
has an opposite effect for fluid particles heavier and lighter than the
surrounding fluid. The case of vorticity in two-dimensional turbulence is also of interest.
Indeed statistical mechanics of two-dimensional turbulence provides predictions for the final
flow organisation depending on an initial distribution of vorticity values\cite{robert,miller}. This theory can be used as a starting point
for non-equilibrium transport models \cite{robert92,kazantsev}, expressed in terms of the local probability density function of vorticity.
Turbulent cascade however modifies this probability distribution in the presence of a small viscosity,
 with a dissipation of its fluctuations. In two-dimensional turbulence, this effect leads to 
a modification of the equilibrium state resulting from turbulent mixing\cite{brands}. We propose here a
  simple model for this cascade effect, which could be combined with transport equations in the presence
  of spatial gradients.

Let us consider a scalar field $\sigma(\mathbf{r},t)$ transported and conserved by
 the divergenceless turbulent motion of fluid parcels. The effect of molecular diffusion is expected
 to smooth out the scalar field at the Batchelor's diffusive cut-off scale $r_d$. We however prefer to
 consider the case of a purely advected scalar, with no diffusion, and introduce a local  average at
  a given scale $l$, $\sigma_l(\mathbf{r},t)= \int
 G_l(\mathbf{r}-\mathbf{r'})\sigma(\mathbf{r'},t) d \mathbf{r'}$, obtained with a linear filtering
 operator $G_l$. This coarse-grained description corresponds to a finite measurement resolution at a scale $l$ larger than $r_d$.
  The fine-grained probability distribution of the scalar is preserved in time as the scalar value
  and the volume of each fluid parcel is conserved in the absence of diffusion. However the coarse-grained probability
   distribution function (pdf) $\rho_l(\sigma,t)$ is not preserved because fluctuations are transferred
  to scales smaller than the cut-off $l$.

The problem is then to find a time evolution equation for this pdf.
The result should of course depend on the properties of the turbulent field. In the usual Kolmogoroff regime,
the dissipation of scalar variance is equal to its cascade flux, independent of the cut-off scale.
This flux  is set by the energy and integral scale of the turbulence, which are described by transport
 equations in empirical turbulence models (like k-epsilon). 

Instead of a Kolmogoroff cascade, we shall here consider a random but smooth and persistent straining motion, in which the velocity difference
 $\mathbf{v}(\mathbf{x}+\mathbf{r},t)-\mathbf{v}(\mathbf{x},t)$ is a linear function of the separation $\mathbf{r}$ \cite{batchelor}. This hypothesis holds whenever the
kinetic energy spectrum is steeper than $k^{-3}$. It could be an appropriate model for vorticity in two-dimensional turbulence.

The straining is charaterized by
the symmetric part of the strain tensor :
$\mathbf{\Sigma}_{ij}=\frac{1}{2}(\partial_j u_i +\partial_i
u_j)$, while the antisymmetric part accounts for the rotation of
a fluid element. As a consequence of the fluid incompressibility, the trace of
$\mathbf{\Sigma}$ is zero and a basis exists
on which this matrix is diagonal with two opposite eigenvalues
 $\pm s(t)$. The axes of an initial spherical blob
corresponding to positive or negative eigenvalues will respectively 
grow or decrease. We assume that the
angle between the positive eigendirection of $\mathbf{\Sigma}$ and
the iso-scalar lines evolves slowly compared to the straining rate $s(t)$,
i.e. the time for the eigendirections to rotate by $\pi/2$ is longer than
the time for the scalar patterns to be strained from the integral scale $L$ to the filtering scale $l$.
With this hypothesis, our problem becomes locally one-dimensional: the fluid is
composed of adjacent sheets of fluid uncorrelated with each other in
one direction (which can be tilted with respect to the eigendirection of
the strain matrix \cite{lapeyre}).

After the time $\Delta t_{1/2}$ needed for the width of a strip to be divided by 2, the
scalar field filtered at scale $l$ becomes the average of two realisations of the field at the previous time. The probabilities of scalar values in adjacent strips can be assumed independent, as they result of the straining of regions which were initially
far apart, at a distance beyond the integral scale of the scalar. Thus the new probability distribution is the self-convolution of the previous
one, describing the sum of the independent random variables, followed by a contraction by a factor 2:
 $\rho_l(\sigma,t+\Delta t_{1/2})=2\int \rho_l(\sigma',t)\rho_l(\sigma-\sigma',t)d\sigma'$.

The convolution can be transformed in a product of the Fourier transform of the pdf (characteristic function).
If the scalar $\sigma$ has only positive values (which can be generally obtained by changing $\sigma$ in $\sigma-min(\sigma)$),
it is more convenient to use the Laplace Transform
 $\widehat{\rho_l}(\kappa)=\int \rho_l (\sigma) e^{-\kappa \sigma} d \sigma$. The Laplace transform of the previous self-convolution relationship
 leads to $\widehat{\rho_l}(2\kappa,t+\Delta t_{1/2}) = [ \widehat{\rho_l} (\kappa,t)]^2$. Similarly, calling $\Delta t_{\frac{1}{n}}$ the time to divide the
thickness of a sheet of fluid by a factor $n$, the pdf at time $t+\Delta
t_{\frac{1}{n}}$ will be an $n^{th}$-selfconvolution. In the spectral
representation, this becomes the product of $n$ identical characteristic functions \cite{feller}:
\begin{equation}
\widehat{\rho_l}(n \kappa,t+\Delta t_{1/n}) = [\widehat{\rho_l}
(\kappa,t)]^n.\label{Eq:discret}
\end{equation}

In order to get a differential equation in time, we now take the limit  $n=1 +\epsilon$ with $\epsilon=s(t)dt$
small in (\ref{Eq:discret}). This yields: $\widehat{\rho_l}(\kappa+\epsilon \kappa,t+dt)=[\widehat{\rho_l}(\kappa,t)]^{1+\epsilon}$ . Taking
 the limit $\epsilon
 \rightarrowtail 0$, we can express ${\rho_l}(\kappa+\epsilon \kappa,t+dt)$ in terms of partial derivatives with respect to $t$ and $\kappa$:
\begin{equation}
\partial_t  \widehat{\rho_l} = s(t) [\widehat{\rho_l} \ln \widehat{\rho_l} -\kappa  \partial_{\kappa} \widehat{\rho_l} ] \label{Eq:etirement_pur}
\end{equation}
One can check that the normalisation $\int \rho_l(\sigma,t) d\sigma=\widehat{\rho_l}(0,t)=1$ is preserved in time by this equation. The mean
scalar value $\int \sigma \rho_l(\sigma,t) d\sigma=-\partial_\kappa \widehat{\rho_l}(0,t)$ is also conserved (if it is initially defined).
Note that if $\widehat{\rho_l}$ were standing for a Fourier instead of a Laplace transform, $\widehat{\rho_l}(\kappa)=\int \rho_l (\sigma) e^{-i\kappa \sigma} d \sigma$, the same
equation would be obtained for the norm of $\widehat{\rho_l}$, with an additional equation for its phase $\phi$: $\partial_t \phi+s(t)\kappa\partial_\kappa\phi=s(t)\phi$.

The right hand term of equation (\ref{Eq:etirement_pur}) describes the effect of strain on the pdf of the scalar, which could be used with additional
terms expressing scalar generation or spatial transport. Equation (\ref{Eq:etirement_pur}) itself can be analytically solved (using the method of characteristics). For that purpose, we define
the integral:
\begin{equation}
f(t) = \exp  (\int_0^t s(t')dt' )\label{eq:fdef}
\end{equation}
(such that $s(t)=f'(t)/f(t)$). We can easily show that a strip with initial width $R(0)$ reaches a width $R(t)=R(0)/f(t)$ at time $t$, so that $f(t)$ is the
reduction factor in the straining process. We can check that the result
\begin{equation}
 \widehat{\rho_l}(\kappa,t)={\bigg [\widehat{\rho_l}\bigg (\frac{\kappa}{f(t)},0\bigg )\bigg ]}^{f(t)} \label{eq:solu}
 \end{equation}
is solution of (\ref{Eq:etirement_pur}).
When $f(t)$ is an integer, we recover the expression (\ref{Eq:discret}) for the effect of $n$ self-convolutions, in agreement with our initial assumption.

The equation (\ref{Eq:etirement_pur}) and its solution (\ref{eq:solu}) describe the process of convergence to a Gaussian stated by the central limit theorem: at
 times goes on, scalar fluctuations
initially extending over more and more area become packed by the straining effect below the filtering scale.
This can be be checked by the convergence to zero of all the cumulants beyond the second order one. The $m^{th}$ cumulant, defined as ${\langle\sigma^m\rangle}_c(t)=
{(-\partial_{\kappa})}^m \ln (\rho_l(\kappa,t)) \big|_{\kappa=0}$, is readily obtained from (\ref{eq:solu}),
\begin{equation}
{\langle\sigma^m\rangle}_c (t)=\frac{{\langle\sigma^m\rangle}_c(0)}{{[f(t)]}^{m-1}} \label{Eq:cumulants}
\end{equation}
The cumulant of order 2, equal to the variance, ${\langle\sigma^2\rangle}_c={\langle\sigma^2\rangle}-{\langle\sigma\rangle}^2$, decays
as $1/f(t)$. This expresses the decay of the scalar variance by the cascade through the filtering
scale. The relative value of the higher order cumulants is expressed as ${\langle\sigma^m\rangle}_c/{\langle\sigma^2\rangle}_c^{m/2}\sim f(t)^{-(m/2-1)}$, so it decays in time, approaching
a Gaussian, for which the cumulants with order larger than $2$ are strictly equal to $0$. Note however that 
if the first or second cumulants are not defined at time $t=0$, the expression (\ref{eq:solu}) converges to a Levy distribution \cite{feller}, another form of stable pdf, although the result (\ref{Eq:cumulants}) for the cumulants would not be applicable.

Equation (\ref{Eq:etirement_pur}) can be extented to the three dimensional case. The symmetric part of the deformation tensor may
have \begin{itemize}
\item one negative eignevalue $s(t)$: the problem remains one dimensionnal,
but the fluid is now seen as a succession of adjacent
iso-scalar planes 
\item two negative eignevalues $s_1(t),s_2(t)$ : filaments are formed instead of sheets and the above description remains  valid provided that $s(t)=s_1(t)+s_2(t)$.
\end{itemize}

In the case of a scalar cascade in usual isotropic turbulence, our hypothesis of a uniform straining rate does not apply, but our approach can be still used in a more empirical way. We have seen indeed that the relative rate of decay of the scalar variance is equal to $s(t)$, and this should be equal to the
flux of scalar variance in the Kolmogoroff cascade, independent of the cut-off scale. Then the model equation (\ref{Eq:etirement_pur}) can be applied to determine the evolution
of the whole pdf. This approach ignores the fluctuations of $s(t)$, which are known to generate internal intermittence\cite{falkovich}. Nevertheless, this can be a good model if cascade is
of limited extent in wave numbers, or if larger sources of intermittence come from spatial gradients.

\begin{figure}
\centering
\includegraphics[width=85mm]{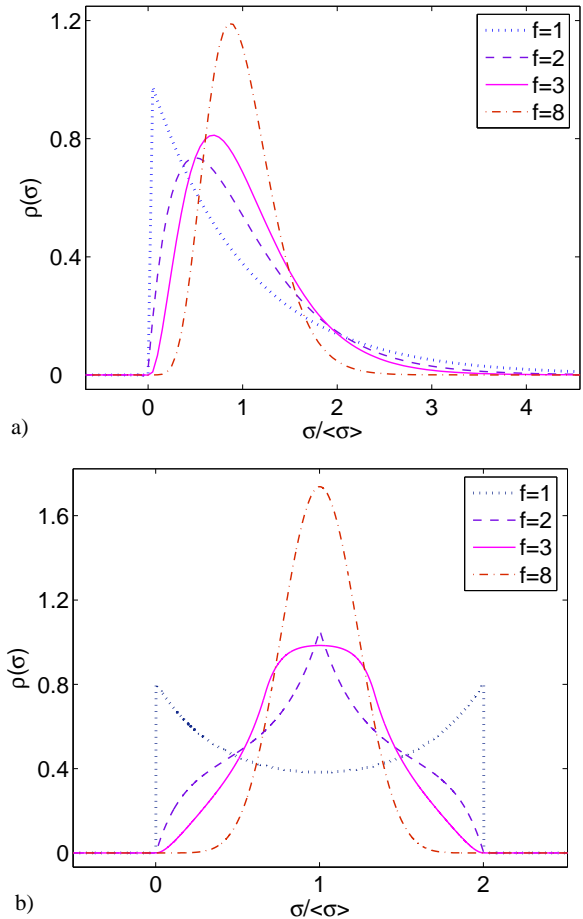}
\caption{ \label{Fig:illu} Evolution of the shape of the scalar
  distribution  as a sel-convolution process when the initial condition is 
 a) strongly asymmetric (a decreasing exponential)
b) symmetric}
\end{figure}

The evolution of the pdf has been studied by \cite{pope} from a numerical computation of isotropic turbulence,
in which a passive scalar is introduced. The initial
pdf was made of two symmetric peaks, and the convergence to a Gaussian pdf with decreasing variance was numerically observed.
The corresponding result from our model is
represented in figure (\ref{Fig:illu}-b). The time scale is arbitrary, but $s(t)$ could be adjusted to fit the observed decay of the scalar variance. Then
the higher order cumulants can be calculated from the model.

As a second example, we consider experiments performed by locally introducing a dye in a stirred flow 
\cite{villermaux}. In that case the initial concentration is $0$ everywhere except in the dye streaks, so the corresponding pdf is very skewed. Later in time it progressively tends to a Gaussian as stirring proceeds. Villermaux and Duplat \cite{villermaux} have provided a quantitative model of this evolution as an
 aggregation process of streaks of scalar, which leads to the following 
kinetic equation : 
\begin{equation}
\partial_t  \widehat{\rho} = s(t) \big[ f(t) [{\widehat{\rho}}^{1+1/f}-\widehat{\rho}]-\kappa  \partial_{\kappa} \widehat{\rho} \big] \label{Eq:villermaux}
\end{equation}
where $f(t)$ is the integrated strain defined by relation (\ref{eq:fdef}) like in our model. The first term accounts for the formation of scalar sheets, while the second one describes the decay of the concentration by the competing effects of strain and diffusion in scalar
streaks. The solution of (\ref{Eq:villermaux}) approaches  a sequence of gamma pdf at large time, $\gamma(\sigma/\langle \sigma \rangle)= \frac{f^f}{\Gamma(f)}\frac{\sigma^{f-1}}{{\langle \sigma \rangle}^{f-1}} e^{-f \sigma/\langle \sigma \rangle} $, whose characteristic function is
\begin{equation}
\widehat{\gamma}(\kappa) = {\bigg( \frac{1}{1+\langle
\sigma \rangle \kappa/f(t)}\bigg)}^{f(t)}\label{Eq:gammapdf}
\end{equation} 
The average concentration $\langle \sigma \rangle$ is a constant, while the exponent $f(t)$ increases in time. The pdf is very skewed at the beginning and becomes more and more symmetric and narrow at time goes on.
The solution (\ref{Eq:gammapdf}) has been found to be in good agreement with the experimental results obtained by introducing dye either in a steady stirring motion or in a turbulent pipe flow.

Note that (\ref{Eq:gammapdf}) is in the form (\ref{eq:solu}), so that it is also a solution of our dynamical equation (\ref{Eq:etirement_pur}). Therefore our model can also account for the experimental results. However in our case, it corresponds to a particular initial condition, an exponential pdf (see figure 1-a). Such an initial condition could result from non-homogeneous processes occuring near the dye injector. Other forms can be obtained from different initial conditions, although with the same qualitative behavior. By contrast, Villermaux and Duplat found that the family of gamma pdf is an attractive solution \cite{villermaux} , so that it can be approached for a wider class of initial conditions. One can easily check that the dynamical equation
(\ref{Eq:villermaux}) becomes identical to (\ref{Eq:etirement_pur}) for long times, $f \to \infty$. The change of the pdf in our model (\ref{Eq:etirement_pur}) depends only on the straining rate $s(t)$, while in (\ref{Eq:villermaux}) it keeps track of the previous history through $f(t)$, which may not be suitable in the presence of spatial fluxes. Another difference with the model of Villermaux and Duplat \cite{villermaux} is that we consider concentration averaged on a small domain  (in the absence of diffusion) instead of pointwise concentration. Distinguishing between the two models would require careful analysis of the experimental data. 

The effect of a turbulent diffusion in the presence of a mean scalar gradient has been studied by \cite{pumir}, who show that the pdf then develops exponential tails. This provides therefore a good rational for an initialisation of our cascade model by an exponential, leading then to solutions close to gamma pdf. In a steady regime sustained by a scalar gradient, the index $f$ would then depends on the ratio of the cascade effect to the spatial fluxes, instead of time.  

Gamma pdf have been also proposed as a fit of density increments at meter scale in the ocean \cite{pinkel}. Such distributions have been reproduced in numerical computations of internal wave breaking in a stably stratified fluid \cite{bouruet}. This was modeled as 
the effect of random steps in density due to mixing processes. However, such steps are not observed. Our approach could provide another justification. Indeed, a simple hypothesis of statistical equilibrium in the gravity field leads to an exponential pdf depending on density $\sigma$ and vertical height $z$. This would provide the Gamma pdf through the self-convolution process. However the link between the statistics of this coarse-grained concentration at scale $l$ and those of vertical density increments at scale $l$ would need to be claridied.  

To conclude, the main result of this letter is the continuous process (\ref{Eq:etirement_pur}) that accounts for the temporal
evolution of the coarse-grained scalar pdf : 
the probe of width $l$ ``sees'' structures coming from  larger and larger scales, which implies the self-convolution of the pdf. 
. The efficiency of the fluid motion to drive the scalar pdf to
a form given by the central limit theorem depends of the rate of
strain $s(t)$. It does not take into account
intermittency by assuming that $s(t)$ is homogeneous in space.
 Spatial gradients could in some cases be the main raison of non-gaussianity of the scalar distribution. 
In that respect, a fuller study using this result
in the context of statistical mechanics of the mixing of a stably
stratified fluid is in preparation and will be reported elsewhere.

\bibliographystyle{apsrev}
\bibliography{mixing}

\end{document}